\documentclass[12Pt]{article} 
\usepackage{amssymb,euscript,latexsym,amsmath} 
\usepackage{graphicx} 
\usepackage{epstopdf}
\usepackage{bm} 
\usepackage{enumerate} 
\usepackage{color} 
\usepackage{setspace} 
\usepackage{subfigure} 
\usepackage{hyperref}

\begin{document}

\title{Insights into symmetric and asymmetric vortex mergers using the core growth model} 
\author{Fangxu Jing, Eva Kanso and Paul K. Newton}

\maketitle 
\begin{abstract}
We revisit the two vortex merger problem (symmetric and asymmetric) for the Navier-Stokes equations using the core growth model for vorticity evolution coupled with the passive particle field and an appropriately chosen time-dependent rotating reference frame. Using the combined tools of analyzing the topology of the streamline patterns along with careful tracking of passive fields, we highlight the key features of the stages of evolution of vortex merger, pinpointing deficiencies in the low-dimensional model with respect to similar experimental/numerical studies. The model, however, reveals a far richer and delicate sequence of topological bifurcations than has previously been discussed in the literature for this problem, and at the same time points the way towards a method of improving the model. 
\end{abstract}

\section{Introduction}

\subsection{Background}
Consider an incompressible fluid in an unbounded two-dimensional domain $\mathbb{R}^2$. The fluid motion is governed by Navier-Stokes equations, written in terms of the vorticity $\omega(\mathbf{x},t)$ as follows 
\begin{equation}
	\frac{\partial\omega}{\partial t} = -\mathbf{u}\cdot\nabla\omega + \nu\Delta\omega \,, \label{eq:n-s} 
\end{equation}
where $\nu$ is the kinematic viscosity taken to be constant, $\mathbf{u}(\mathbf{x},t)$ is the fluid velocity as a function of time $t$ and the position vector $\mathbf{x}$. Both $\mathbf{x}$ and $\mathbf{u}$ can be expressed in inertial frame $\{\mathbf{e}_i\}_{i=1,2,3}$ where $(\mathbf{e}_1,\mathbf{e}_2)$ span the plane of motion. For simplicity, we denote the nonzero components $\mathbf{x} = (x,y)$ and $\mathbf{u} = (u_x,u_y)$. The vorticity vector $\boldsymbol{\omega} = \nabla \times \mathbf{u}$ which is always perpendicular to the plane of motion is therefore denoted by its only non-zero component $\omega$ in the $\mathbf{e}_3$-direction. The velocity $\mathbf{u}$ and vorticity $\omega$ are related via the 2D Biot-Savart law 
\begin{equation}
	\mathbf{u}(\mathbf{x},t) = \frac{1}{2\pi} \int_{\mathbb{R}^2} \frac{(\mathbf{x} - \tilde{\mathbf{x}})^{\perp}}{\|\mathbf{x} - \tilde{\mathbf{x}}\|^2} \, \omega(\tilde{\mathbf{x}},t) \, \text{d}\tilde{\mathbf{x}}\,, \label{eq:b-s} 
\end{equation}
where $\tilde{\mathbf{x}}$ is an integration variable and $\mathbf{x}^{\perp} = ( -y , x )$. The solution of the system of equations~\eqref{eq:n-s} and~\eqref{eq:b-s} depends on the initial condition $\omega(\mathbf{x}, 0)$. One solution of particular interest in this work is the Lamb-Oseen solution corresponding to a point vortex initial condition $\omega(\mathbf{x}, 0) = \Gamma \delta(\mathbf{x})$, centered at the origin with circulation $\Gamma$. Traditionally, the problem can be expressed compactly in complex notation with position variable $\mathbf{z} = x + i y$. The Lamb-Oseen vorticity and velocity fields are given by 
\begin{equation}
	\omega(\mathbf{z},t) = \frac{\Gamma}{4\pi\nu t}\exp \left(-\frac{\|\mathbf{z}\|^2}{4\nu t}\right) \,, \qquad \dot{\mathbf{z}}^* = u_x - i u_y = \frac{\Gamma}{2 \pi i} \frac{1}{\mathbf{z}} \left[1 - \exp\left(-\frac{\|\mathbf{z}\|^2}{4 \nu t}\right)\right] \,, \label{eq:singlegaussian} 
\end{equation}
where the notation $\dot{()} = {\rm d}()/{\rm d}t$ refers to the time derivative and the notation $\mathbf{z}^*$ refers to the complex conjugate of $\mathbf{z}$. The vorticity field is initially concentrated at the origin then diffuses axisymmetrically as a Gaussian distribution. The spreading of the vorticity can be quantified by the vortex core defined as $a = \sqrt{4 \nu t} \equiv \sqrt{4\tau}$. 

For more complicated initial condition, exact solutions of~\eqref{eq:n-s} and~\eqref{eq:b-s} with ``growing cores'' are not analytically available in general. We are particularly interested in the viscous evolution of a class of initial vorticity fields $\omega(\mathbf{z},0) = \sum_{\alpha=1}^N \Gamma_\alpha\delta(\mathbf{z} - \mathbf{z}_\alpha)$ consisting of the superposition of $N$ point vortices. Gallay \& Wayne recently proved that the Lamb-Oseen solution is an asymptotically stable attracting solution for all integrable initial vorticity fields, see~\cite{GaWa2005a}, of which the $N$ point vortices is a special case. Our goal in this paper is to elucidate the dynamics that takes place as a given system evolves to the Lamb-Oseen solution. We use a familiar {\em core-growth} model (see our previous works~\cite{JiKaNe2010a, Ji2011a}) on the two-vortex merger problem in an appropriate rotating reference frame with the knowledge that this model is not exact, but is nonetheless extremely useful when carefully implemented on a problem where the intermediate evolution has been carefully described both experimentally and numerically in the literature. The comparisons will allow us to pinpoint the shortcomings of the model in this context and build ``higher fidelity'', but still low-dimensional physically based corrections. 

\begin{figure}
	[!t] 
	\centering 
	\includegraphics[width=0.52\textwidth]{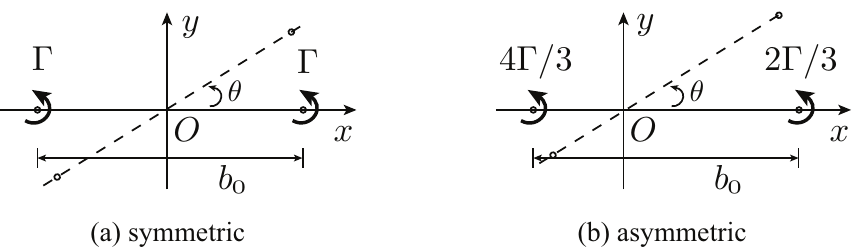}
	\caption{\footnotesize Schematics of (a) symmetric and (b) asymmetric co-rotating vortex pairs.} \label{fig:schematics} 
\end{figure}
We examine the viscous evolution associated with two initial configurations: a symmetric and an asymmetric co-rotating vortex pair. In other words, we consider two initial conditions: (i) two point vortices with equal circulation $\Gamma_L = \Gamma_R = \Gamma$ located at $\mathbf{z}_L(0) = - b_0/2$ and $\mathbf{z}_R(0) = b_0/2$, and (ii) two point vortices with circulation $\Gamma_L = 4 \Gamma/3$ and $\Gamma_R = 2\Gamma/3$ located at $\mathbf{z}_L(0) = - b_0/3$ and $\mathbf{z}_R(0) = 2b_0/3$, see Figure~\ref{fig:schematics}. In both cases, the total circulation is $2\Gamma$ and the separation distance between the two vortices is $b_0$. The inviscid $(\nu=0)$ evolution of these two cases is well understood. Since the induced velocities from the point vortices on each other are always perpendicular to the connecting line, the distance between the vortices remains constant $b_{0}$ for all time, and the vortices are co-rotating with a constant rotation rate $\dot{\theta}$ around the origin 
\begin{equation}
	\dot{\theta} = \frac{\Gamma_L+\Gamma_R}{2\pi b_{0}^2} \,,\label{eq:thetadotpv} 
\end{equation}
all of which is well known and discussed in~\cite{Ne2001a}. The rotation angle $\theta$ can be integrated from~\eqref{eq:thetadotpv}. The position of the left vortex is $\mathbf{z}_L = -(b_0/2) e^{i\theta}$ and the right vortex $\mathbf{z}_R = (b_0/2) e^{i\theta}$ for the symmetric case, whereas $\mathbf{z}_L = -(b_0/3) e^{i\theta}$ and $\mathbf{z}_R = (2b_0/3) e^{i\theta}$ for the asymmetric case. In general, the inviscid evolution of $N$ point vortices is known to exhibit regular, including fixed and relative equilibria, as well as chaotic dynamics depending on the number of vortices, their strengths and initial positions, see~\cite{Ne2001a,Ar1983a,Sa1992a,Ar2010a}.

Our goal in this paper is to describe the dynamical evolution of these co-rotating vortices for non-zero viscosity ($\nu\neq 0)$. Instead of numerically solving the Navier-Stokes equation, we use a simple, analytically-tractable, low-dimensional model. For a comparison between the model and Navier-Stokes equation with a different initial configuration, the reader is referred to~\cite{JiKaNe2010a, Ji2011a}. The model assumes that the vorticity of each initial point vortex spreads axisymmetrically as an isolated Lamb-Oseen vortex, thus modeling the diffusion term $\nu \Delta \omega$ in~\eqref{eq:n-s}, while its center moves according to the local velocity induced by the presence of the other diffusing vortices, thus accounting for the convection term $\mathbf{u}\cdot\nabla \omega$ in~\eqref{eq:n-s}. Stated differently, the model assumes that the vorticity field is a superposition of multiple Lamb-Oseen vortices 
\begin{equation}
	\omega = \sum_{\alpha = 1}^N \frac{\Gamma_{\alpha}}{4 \pi\tau} \ \exp\left(\frac{-\|\mathbf{z}-\mathbf{z}_{\alpha}\|^2}{4 \tau}\right) \,, \label{eq:omegamgm} 
\end{equation}
where $\tau \equiv \nu t$ is time scaled by viscosity. The velocity field is computed by substituting \eqref{eq:omegamgm} into \eqref{eq:b-s}, 
\begin{equation}
	\dot{\mathbf{z}}^* = \sum_{\alpha = 1}^N \frac{\Gamma_{\alpha}}{2 \pi i (\mathbf{z} - \mathbf{z}_{\alpha})} \left[1 - \exp\left(\frac{-\|\mathbf{z}-\mathbf{z}_{\alpha}\|^2}{4 \tau}\right)\right] \,. \label{eq:velmgm} 
\end{equation}
And the velocity of the center $\mathbf{z}_\beta$ of the $\beta^{th}$ vortex is given by subtracting the effect of itself from~\eqref{eq:velmgm}, 
\begin{equation}
	\dot{\mathbf{z}}^*_{\beta} = \sum_{\alpha \neq \beta}^N \frac{\Gamma_{\alpha}}{2 \pi i (\mathbf{z}_{\beta} - \mathbf{z}_{\alpha})} \left[1 - \exp\left(\frac{-\|\mathbf{z}_{\beta}-\mathbf{z}_{\alpha}\|^2}{4 \tau}\right)\right] \,. \label{eq:centervelmgm} 
\end{equation}
The system of equations in~\eqref{eq:omegamgm}, \eqref{eq:velmgm}, and~\eqref{eq:centervelmgm} is referred to as the {\em multi-Gaussian} model, or the core-growth model, discussed as the basis for a numerical scheme, for example, in \cite{CoKo2000a}. It is useful to notice in~\eqref{eq:centervelmgm} that the entire numerator can be thought of as the vortex strength $\Gamma_\alpha (\mathbf{z}_{\alpha}, \mathbf{z}_{\beta}; \tau)$ which is both time-dependent, and depends on the position of each vortex with respect to all of the other positions. Thus it is not surprising that the transient dynamics associated with the model depends on the initial configuration chosen. We will augment this basic model by coupling it with an appropriately chosen passively advected field, discussed further in Section 2.

\subsection{Some relevant literature on vortex merger}

The merging of two co-rotating vortices is an essential phenomenon in transitional and turbulent flows and as a result, has emerged as a ``canonical'' problem in the fluid dynamics literature. It seems to be the ``simplest'' problem whose evolution involves the interplay between the two key terms in the equations of motion~\eqref{eq:n-s} for incompressible flows. First is the {\em convective} term given by $\mathbf{u}\cdot\nabla\omega$ in the vorticity form of the Navier-Stokes equations, second is the {\em viscous} term given by $ \nu\Delta\omega$. Of course both operate simultaneously in evolving the initial configuration, through the transients, to the final Lamb-Oseen state. The convective term is usually thought of as being responsible for aspects of merging such as filamentation and stretching (inviscid  aspects), while the viscous term is of course responsible for the ultimate diffusive merging. Previous works on vortex merging can be categorized into symmetric versus asymmetric vortex pairs, or inviscid versus viscous fluid model. The related results are briefly described here in roughly chronological order.

Earlier studies mainly focused on mergers in inviscid fluid, i.e. solving the Euler equations. Saffman \& Szeto~\cite{SaSz1980a} found the critical separation between the symmetric vortex pair by solving for steady equilibrium shapes of the contour of uniform vorticity patches (sometimes also referred to as {\em Rankine} patches). Overman \& Zabusky~\cite{OvZa1982a} confirmed Saffman \& Szeto's result using contour dynamics for the symmetric uniform patches, and determined which vortex ``wins'' for the asymmetric vortex pair (a point vortex and an extended uniform patch). Dritschel~\cite{Dr1985a} used a contour surgery method to study symmetric uniform patches and linked shape stability to the energy of the system. He also, for the first time, showed the importance of separatrices of the streamlines in a co-rotating reference frame for the point vortex model. Melander \textit{et al.}~\cite{MeZaMc1987a, MeZaMc1988a} studied both the symmetric and asymmetric mergers of uniform patches. They juxtaposed the results of a high resolution pseudo-spectral method to solve the Euler equations with a hyper-viscosity term ($\Delta^2 \omega$) and a low order moment model. In the co-rotating frame, they described flow regions bounded by the separatrices of the streamlines, namely, the {\em core regions} around the vortices, an {\em exchange band} that entraps the cores, and the region outside the exchange band. They also noted the ``ghost'' vortices on the vertical axis in the co-rotating frame, which appear due to the eccentric nature of the flow field generated by the vortex pair. Dritschel \& Waugh~\cite{DrWa1992a} used contour dynamics to study asymmetric uniform patches with the same vorticity density, and identified 5 flow regimes of vortex interaction depending on area ratio and separation of vortices: elastic interaction, partial straining-out, complete straining-out, partial merger and complete merger. Yasuda \& Flierl~\cite{YaFl1995a}  studied the same problem, but added one more parameter to vary -- vorticity density ratio, and confirmed the categorization of the 5 flow regimes. Ehrenstein \& Rossi~\cite{EhRo1999a} studied contour dynamics for both symmetric and asymmetric cases, but instead of uniform vorticity patches, they focused on  Gaussian-like patches with compact support. They noticed the critical distance of merging in this case is less than that for uniform patches.

In 2001, Amoretti \textit{et al.}~\cite{AmDuFaPo2001a} compared an electron-plasma experiment and a particle-in-cell numerical simulation of the Euler equations for an asymmetric setup -- a point vortex and an uniform patch. In the same year, Meunier \& Leweke~\cite{MeLe2001a} conducted water tank experiments to study the instability of 3D vortex pairs, which provided remarkable insight to the 2D problem as well. They described the merging of a symmetric pair as 3 stages based on the separation and core size of the vortices (defined by the maximum azimuthal velocity): (i) the separation between vortices remains almost constant and the core size grows like a Gaussian diffusion; (ii) when the core size grows to about 30\% of the separation distance, filamentation becomes much more obvious, and the two vortices merge; (iii) the merged vortices diffuse like a Gaussian again. They also mentioned that a complete merge always occurs in a viscous flow, which is evident from~\cite{GaWa2005a} since a Gaussian is the eventual solution of any vorticity field. Later, Le Diz\`{e}s \& Verga~\cite{LeVe2002a} performed direct numerical simulation of the evolution of initial Gaussian vortices using a spectral method, and analyzed the local eccentricity around vortices before the merging starts. Meunier \textit{et al.}~\cite{MeEhLeRo2002a} compared their water tank experimental result and analytical result (continuation analysis of an equilibrium solution for the Euler equations with initial symmetric Gaussian patches), and obtained an expression for the separation distance as a function of time and Reynolds number.

Cerretelli \& Williamson~\cite{CeWi2003a} conducted an experiment of symmetric vortex merger shed from the trailing edge of a wing in water tank, and analyzed the flow field in the co-rotating frame to explain the physical mechanism of merging. They convincingly identified 4 stages of the merging process (their Figure 4) at Reynolds number 530: (i) the first diffusive stage, (ii) the convective stage, (iii) the second diffusive stage (which was not clearly noticed in Meunier \& Leweke~\cite{MeLe2001a}), and (iv) the merged diffusive stage. They superimposed separatrices of streamlines in a co-rotating frame, as well as vorticity contours,  and noted the clear correlation between them. More importantly, they decomposed the vorticity field into a \textit{symmetric} part (two Gaussians) and an \textit{antisymmetric} part (a vortex quadrupole), and concluded the first diffusive stage is primarily responsible for spreading the initially confined concentrated vorticity profile to an outer {\em recirculation} region of the flow whose velocity field redistributes the vorticity to form two pairs of weaker vortex dipole structures (i.e. a quadrupole) which by its nature pushes the centers of the original stronger vortices together. Then,   the second and merged diffusive stages act to finish job. Meunier \textit{et al.}~\cite{MeLeLe2005a} confirmed Cerretelli \& Williamson's result with a water tank experiment and a spectral method in direct numerical simulation, and also emphasized that the merging time changes with Reynolds number. Velasco Fuentes~\cite{Fu2005a} used a vortex-in-cell method to study uniform symmetric patches in inviscid flow, and concluded that filamentation is not the reason for merging, but merely a result of the process. He also brought up a point that has often been overlooked in the literature: the difference between the Lagrangian and Eulerian points of view in the merging problem, i.e. between trajectories  of passive particles (dye, or weak vorticity) and streamlines -- especially separatrices -- in  a rotating frame. Trieling \textit{et al.}~\cite{TrFuHe2005a} revisited the asymmetric merger problem using contour dynamics in inviscid flow. Instead of focusing on an equilibrium analysis like some of the previous studies, they showed a comparison between the simulation and electron-plasma and water tank experiments during the transient stage of merging. The dye visualization of the water tank experiment for different circulation ratios (their Figure~15) is especially worth mentioning. More recently, Brandt \& Nomura~\cite{BrNo2007a, BrNo2010a} performed high fidelity directly numerical simulation of the Navier-Stokes equations using a pesudo-spectral method for both symmetric and asymmetric mergers, and related the merging process to the strain field generated by the deformed vortices. They once again pointed out the importance of the comparison between the vorticity contours and streamlines in a co-rotating frame.

Although different studies focused on pinpointing the distinct roles played by the convective and diffusive terms, in reality, both terms operate simultaneously. Untangling the ``order'' in which their presence is felt in our mind is not a well defined problem, and is analogous to answering a ``chicken vs. egg'' question. Among the many complications in elucidating these issues separately is the fact that they depend on the Reynolds number associated with the system~\cite{CeWi2003a, MeLeLe2005a}. Our low-dimensional model described in this paper captures well the first diffusive stage, the second diffusive stage, and the merged diffusive stage described in Cerretelli \& Williamson~\cite{CeWi2003a}, but cannot generate quadrupole vorticity structures, hence does not capture the essential feature of the convective merging stage. By this, we do not mean to imply that the model ignores all convective (inviscid) aspects of the transient dynamics, only those that are able to generate the quadrupole moment of vorticity. This is the essential deficiency in using the core growth model to study two-vortex merging, but also points to a low-dimensional way to correct the deficiency which we discuss at the end.

Determination of a merging criterion for the vortex pairs has been the focus of many studies both in inviscid and viscous setups, mostly in the symmetric case~\cite{SaSz1980a,OvZa1982a,MeEhLeRo2002a,BrNo2007a}, and fewer in the asymmetric case~\cite{DrWa1992a,TrFuHe2005a,BrNo2010a}. We study a symmetric case with identical circulations, $\Gamma$, initially assigned to each vortex, and an asymmetric case with unequal circulations $4\Gamma/3$ and $2\Gamma/3$ initially assigned to each vortex. Both cases have the same total circulation of $2\Gamma$ and are shown in Figure \ref{fig:schematics}. Both cases will approach the same asymptotic Lamb-Oseen solution in long time since their total circulation is the same. However, the transient dynamics is rather distinct and interesting, which will be shown in the later sections. For the symmetric case, the onset of merging is usually determined by the ratio between the vortex core size $a$ and the separation between the vortex centers $b$. The critical ratio $(a/b)_{cr}$ has been found to be between 0.23 and 0.31 in the mentioned studies. For the asymmetric case, as stated in several studies mentioned before, whether a universal critical ratio can be found still remains an open question. 

The organization of this paper is as follows: we implement the multi-Gaussian model in the context of the co-rotating pairs in Section~\ref{sec:mergers}. We describe the evolution of vorticity and absolute velocity fields for the symmetric and asymmetric cases in Section~\ref{sec:absolute}. We then show the relative velocity field in Section~\ref{sec:relative}, and focus on the evolution and bifurcation of the instantaneous fixed points and instantaneous separatrices. Since weaker vorticity tends to be ``passively'' advected by stronger more concentrated vorticity in the full equations of motion (i.e. {\em one-way} coupling of the vorticity field), a key diagnostic tool for us is the tracking of passive tracers that are seeded judiciously in the flow field. The evolution of these passive patches are presented and discussed in Section~\ref{sec:tracer}. Finally, in Section~\ref{sec:discussion} we make connections between the results obtained from the multi-Gaussian model and the previous studies and point to ways in which the low-dimensional core-growth model used in this paper can be augmented with the quadrupole terms to increase its fidelity to the full Navier-Stokes system.

\section{Viscous evolution of symmetric and asymmetric mergers}\label{sec:mergers} 
\begin{figure}
	[!tb] 
	\centering 
	\includegraphics[width=0.62\textwidth]{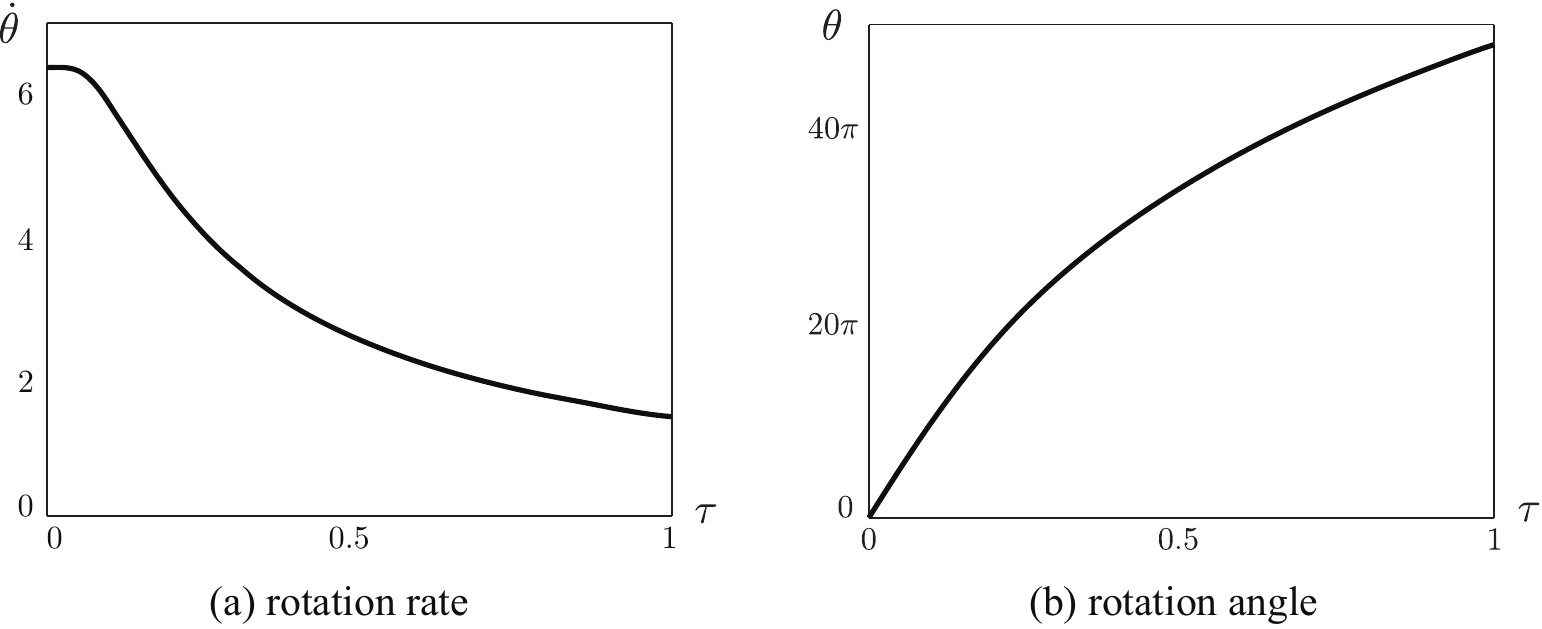}
	\caption{\footnotesize (a) Rotation rate $\dot{\theta}$ and (b) rotation angle $\theta$ for the co-rotating vortex pair, $Re = 1000$.} \label{fig:theta} 
\end{figure}

We examine the dynamics of the examples given in Figure~\ref{fig:schematics} using the multi-Gaussian model~\eqref{eq:omegamgm},~\eqref{eq:velmgm} and~\eqref{eq:centervelmgm}. We nondimensionalize the problem by length scale $L = b_{0}$, and time scale $T = 2\pi^{2} b_{0}^2/\Gamma$ which is the inviscid turnover time. The Reynolds number in viscous vortex dynamics is defined as $Re = \Gamma/\nu$. We show $Re = 1000$ as an example. All variables are dimensionless hereafter.

\subsection{Vorticity and absolute velocity fields}\label{sec:absolute} 

\paragraph{Symmetric case} For the symmetric co-rotating pair, the vorticity field in~\eqref{eq:omegamgm} takes the form: 
\begin{equation}
	\omega = \frac{\Gamma}{4 \pi\tau}\left[ \exp\left(\frac{-\|\mathbf{z}-\mathbf{z}_L\|^2}{4 \tau}\right) + \exp\left(\frac{-\|\mathbf{z}-\mathbf{z}_R\|^2}{4 \tau}\right)\right] \,. \label{eq:symmomega} 
\end{equation}
Due to symmetry, similar to the inviscid case, the vortex centers are also in relative equilibrium, rotating around the origin. However, the rotation rate is no longer a constant, instead it is given by 
\begin{equation}
	\dot{\theta} = \frac{\Gamma}{\pi b_{0}^2} \left[1 - \exp\left(\frac{-b_{0}^2}{4\tau}\right)\right] \,. \label{eq:thetadot} 
\end{equation}
\begin{figure} 
	[!t] 
	\centering 
	\includegraphics[width=0.95\textwidth]{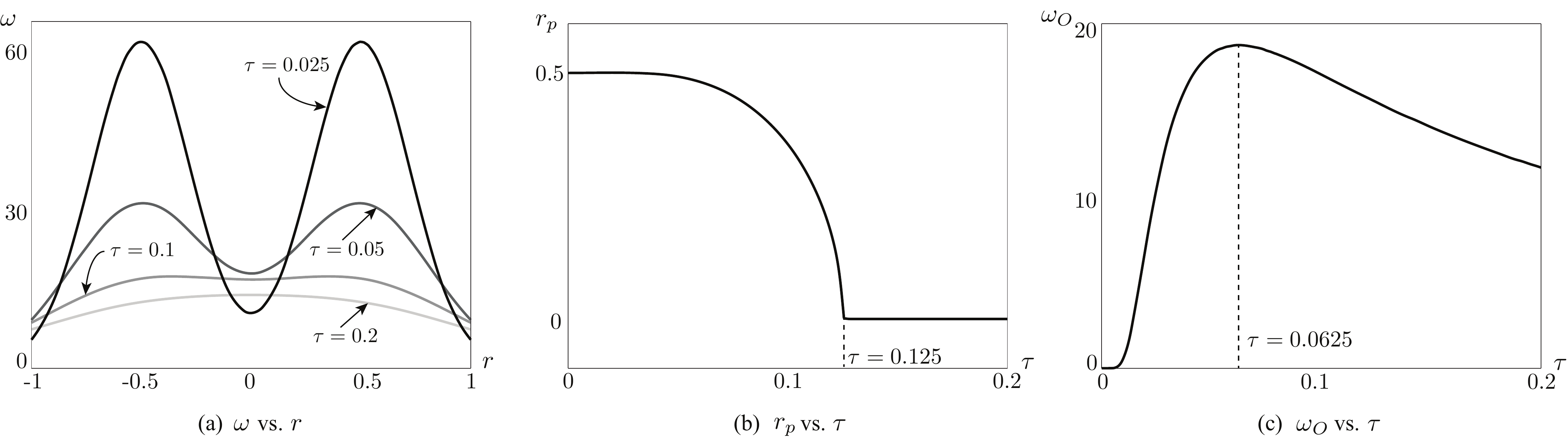}
	\caption {\footnotesize Symmetric vortex pair with $Re = 1000$: (a) Vorticity along the connecting line. The two vorticity peaks eventually decay and merge to a single peak at the origin. (b) Distance between the vorticity peaks and origin as function of time. The peaks merge at time $\tau = 0.125$. (c) Vorticity evolution at the origin. The maximum vorticity occurs at $\tau = 0.0625$.} \label{fig:symmomega}
\end{figure}
\begin{figure}
	[!tb] 
	\centering 
	\includegraphics[width=0.9\textwidth]{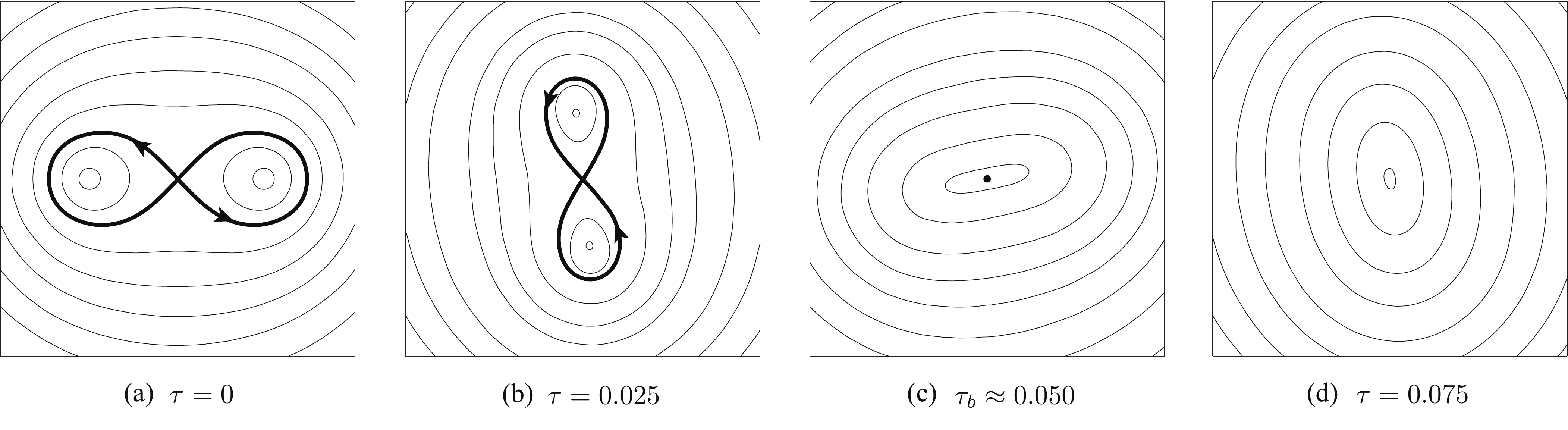}
	\caption{\footnotesize Symmetric vortex pair with $Re = 1000$: Streamlines of the absolute velocity field in inertial frame at $\tau = 0, 0.025, 0.050$ and $0.075$. Each plot shows a $[-1, 1]\times[-1, 1]$ window in $(x,y)$ plane.} \label{fig:symminertial} 
\end{figure}
Integrating $\dot{\theta}$ in time, one has the orientation angle $\theta$ as a function of time, 
\begin{equation}
	\theta = \frac{\Gamma Re}{\pi b_{0}^2} \left[\tau - \exp\left(\frac{-b_{0}^{2}}{4\tau}\right) + \frac{b_{0}^2}{4}\text{Ei}\left(\frac{-b_{0}^2}{4\tau}\right)\right]\,, \label{eq:theta} 
\end{equation}
in which the Euler function is defined as $\text{Ei}(x) = \int_{-\infty}^x \exp(\tilde{x})/\tilde{x} \ \text{d}\tilde{x}$ in the sense of principle value. Figure~\ref{fig:theta} shows $\dot{\theta}$ and $\theta$ as functions of the scaled time $\tau$. Initially, $\dot{\theta}(0)$ is the same value as the inviscid case~\eqref{eq:thetadotpv}, then gradually decays to zero as $\tau\rightarrow\infty$. The positions of vortex centers are given by 
\begin{equation}
	\mathbf{z}_{R} = -\mathbf{z}_{L} = \frac{b_{0}}{2} e^{i\theta} \,.\label{eq:symmzrzl} 
\end{equation}
The vorticity along the line that connects the vortex centers is plotted at various instants in Figure~\ref{fig:symmomega}(a). The distance between the symmetric vorticity peaks and origin is denoted $r_{p}$. At $r_{p}$, one has, $ 
\partial\omega/ 
\partial r = 0$ and $ 
\partial^{2}\omega/ 
\partial r^{2} \leq 0$, $r$ being the distance from the origin. The result is plotted in Figure~\ref{fig:symmomega}(b). The turning point is obtained when the second derivative is equal to 0, when $\tau = 0.125$. As mentioned in previous studies (e.g.~\cite{CeWi2003a}), even if the location of vortex centers remain fixed, due to diffusion, the vorticity peaks will merge in finite time. The vorticity at the origin is given by $\omega_O= (\Gamma/2\pi\tau) \exp(-b_{0}^{2}/16\tau)$, and is plotted in Figure~\ref{fig:symmomega}(c). The maximum vorticity at the origin occurs when $\tau = 0.0625$, which is obtained by solving $\tau$ such that $ 
\partial\omega_{O}/ 
\partial\tau = 0$. 

The absolute velocity field is given by substituting~\eqref{eq:symmzrzl} into~\eqref{eq:velmgm}, which gives 
\begin{equation}
	\dot{\mathbf{z}}^{*} = \frac{\Gamma}{2\pi i} \left\{\frac{1}{\mathbf{z}-\mathbf{z}_{L}}\left[1 - \exp\left(\frac{-\|\mathbf{z}-\mathbf{z}_{L}\|^{2}}{4\tau}\right)\right] + \frac{1}{\mathbf{z}-\mathbf{z}_{R}}\left[1 - \exp\left(\frac{-\|\mathbf{z}-\mathbf{z}_{R}\|^{2}}{4\tau}\right)\right]\right\} \,. \label{eq:symmvel} 
\end{equation}
The streamlines associated with the velocity field are depicted in Figure~\ref{fig:symminertial}. Two main features can be observed: (i) the flow field eventually evolves into a single Gaussian vortex as predicted; and (ii) while the structure is evolving, it is also rotating unsteadily with rotation rate given by~\eqref{eq:thetadot}. The \emph{instantaneous stagnation points} in the flow field can be obtained by setting the right hand side of~\eqref{eq:symmvel} to be zero, and their elliptic or hyperbolic characters can be determined by analyzing the eigenvalue problem associated with the linearized system around the instantaneous stagnation points (see~\cite{JiKaNe2010a} for a detailed analysis of the collinear state, and~\cite{Ji2011a} for some details of this work). Initially, two elliptic points coincide with the two vortices and a hyperbolic point at the origin. As $\tau > 0$, the elliptic points begin to move towards the origin, then at a finite bifurcation time $\tau_{b} \approx 0.050$, they collapse at the origin, transforming the origin into an elliptic point. The topology of streamlines changes accordingly. Among the streamlines, the ones that pass through the instantaneous hyperbolic point are called \emph{instantaneous separatrices}, i.e. the thick lines in Figure~\ref{fig:symminertial}, with arrows representing the directions of the flow. Initially, the separatrix is in $\infty$-shape centered at the origin, and the flow field is divided into three regions. As time evolves, the separatrix shrinks towards the origin while the elliptic points converge. Eventually, at $\tau_b$, the separatrix collapses at the origin, and the flow field is characterized by only one type of streamlines. Note that, contrary to a steady state flow field, a time-dependent separatrix does \emph{not} constitute barriers to fluid motion and fluid particles typically move across this separatrix. We would like to point out that the bifurcation time of the velocity field ($\tau_b \approx 0.050$) does not coincide with the time that vorticity peaks collapse at the origin ($\tau = 0.125$), which indicates that the absolute velocity field does not directly relate to the vorticity field.

\begin{figure}
	[!t] 
	\centering
	\includegraphics[width=0.95
	\textwidth]{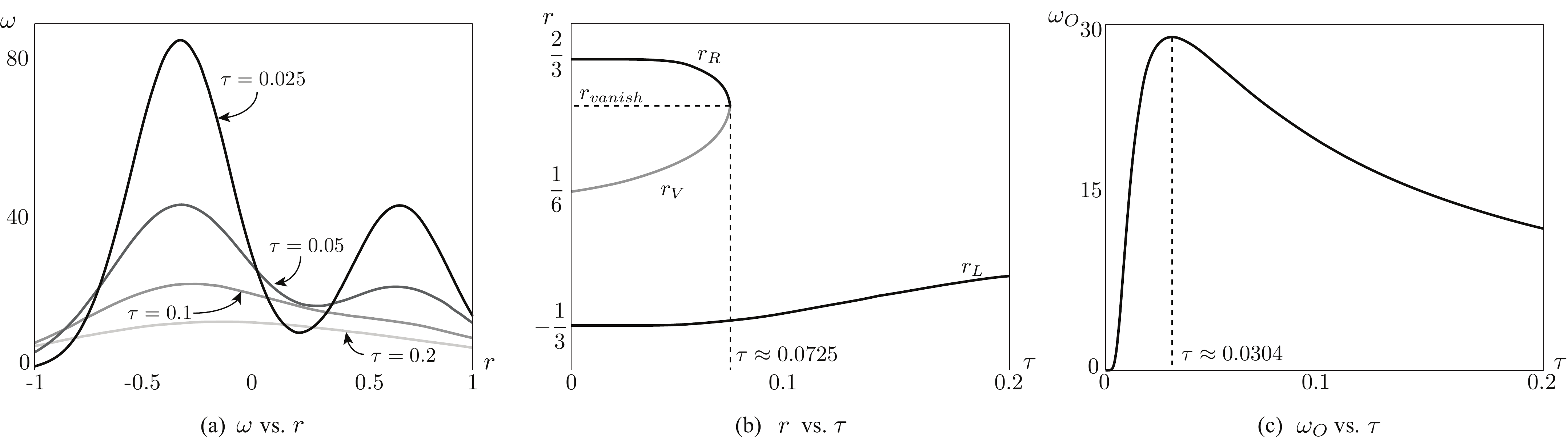}
	\caption {\footnotesize Asymmetric vortex pair with $Re = 1000$: (a) Vorticity along the connecting line. One vorticity peak eventually decay and merge with the valley. (b) Distance between the vorticity peaks (and valley) and origin as function of time. One peak and the valley merge at time $\tau \approx 0.0725$. (c) Vorticity evolution at the origin. The maximum vorticity occurs at $\tau \approx 0.0304$.}\label{fig:asymmomega}
\end{figure}
\begin{figure}
	[!t] 
	\centering 
	\includegraphics[width=0.9\textwidth]{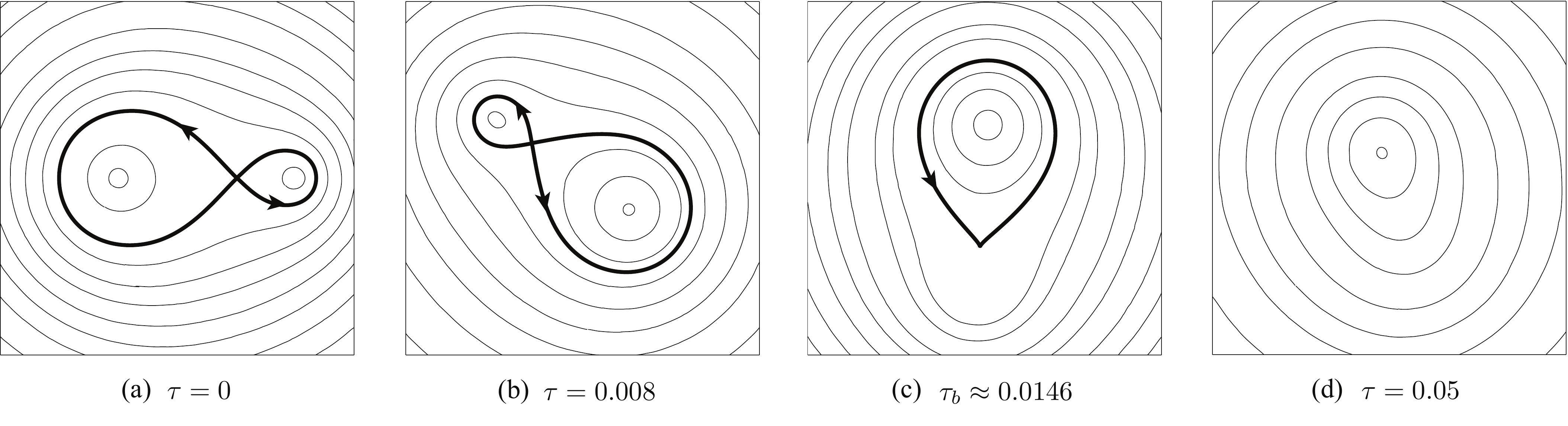}\caption{\footnotesize Asymmetric vortex pair with $Re = 1000$: Streamlines of absolute velocity field in inertial frame at $\tau = 0, 0.008, 0.0146$ and $0.05$. Each plot shows a $[-1, 1]\times[-1, 1]$ window in $(x,y)$ plane.} \label{fig:asymminertial} 
\end{figure}

\paragraph{Asymmetric case}

For the asymmetric co-rotating pair, the vorticity field in~\eqref{eq:omegamgm} takes the form: 
\begin{equation}
	\omega = \frac{\Gamma}{4 \pi\tau}\left[ \frac{4}{3}\exp\left(\frac{-\|\mathbf{z}-\mathbf{z}_L\|^2}{4 \tau}\right) + \frac{2}{3}\exp\left(\frac{-\|\mathbf{z}-\mathbf{z}_R\|^2}{4 \tau}\right)\right] \,. \label{eq:asymmomega} 
\end{equation}
The rotation rate and angle are the same with the symmetric case given in~\eqref{eq:thetadot} and~\eqref{eq:theta}. Indeed, the rotation rate only depends on the \emph{total} circulation of the two vortices and their initial distance, \emph{not} on the strength ratio of the two vortices. The positions of vortex centers are given by 
\begin{equation}
	\mathbf{z}_{L} = -\frac{1}{3} b_{0} e^{i\theta} \,,\quad \mathbf{z}_{R} = \frac{2}{3} b_{0} e^{i\theta}\,.\label{eq:asymmzrzl} 
\end{equation}
The vorticity along the connecting line for the asymmetric case is plotted at various instants in Figure~\ref{fig:asymmomega}(a). The distances between the vorticity peaks and origin are denoted $r_{L}$ and $r_{R}$ for the respective (initial) left and right vortices. The distance between the vorticity valley and origin is denoted as $r_{V}$. The result is plotted in Figure~\ref{fig:asymmomega}(b). Similar to the symmetric case, the vorticity field eventually becomes single peaked. However, instead of the two peaks and one valley all collapse to form the single peak as in the symmetric case, the right peak and valley collapse when $\tau \approx 0.0725$ at $r_{R} = r_{V} = b_0/2$, and annihilate each other, and the left peak remains to be the single peak in the field. The evolution of vorticity at origin is $\omega_{O} = (\Gamma/6\pi\tau) [2 \exp(-b_{0}^{2}/36\tau) + \exp(-b_{0}^{2}/9\tau)]$, it is plotted in Figure~\ref{fig:asymmomega}(c). The maximum vorticity at origin occurs when $\tau \approx 0.0304$.

The streamlines of asymmetric case in inertial frame are depicted in Figure~\ref{fig:asymminertial} at various instants. Initially, the topology of separatrix is the same $\infty$-shape as that of the symmetric case, though the left and right rings are in different sizes. However, at bifurcation time $\tau_{b} \approx 0.0146$, the hyperbolic point collapses with one of the elliptic points and they vanish, only one elliptic point remains. The position of the remaining elliptic point approaches the origin as time goes to infinity. Note that, similar to the symmetric case, the bifurcation time of the velocity field does not coincide with the time that right peak and valley merge in the vorticity field.

\subsection{Relative velocity field}\label{sec:relative} 
\begin{figure}
	[!b] \centering 
	\includegraphics[width=0.4
	\textwidth]{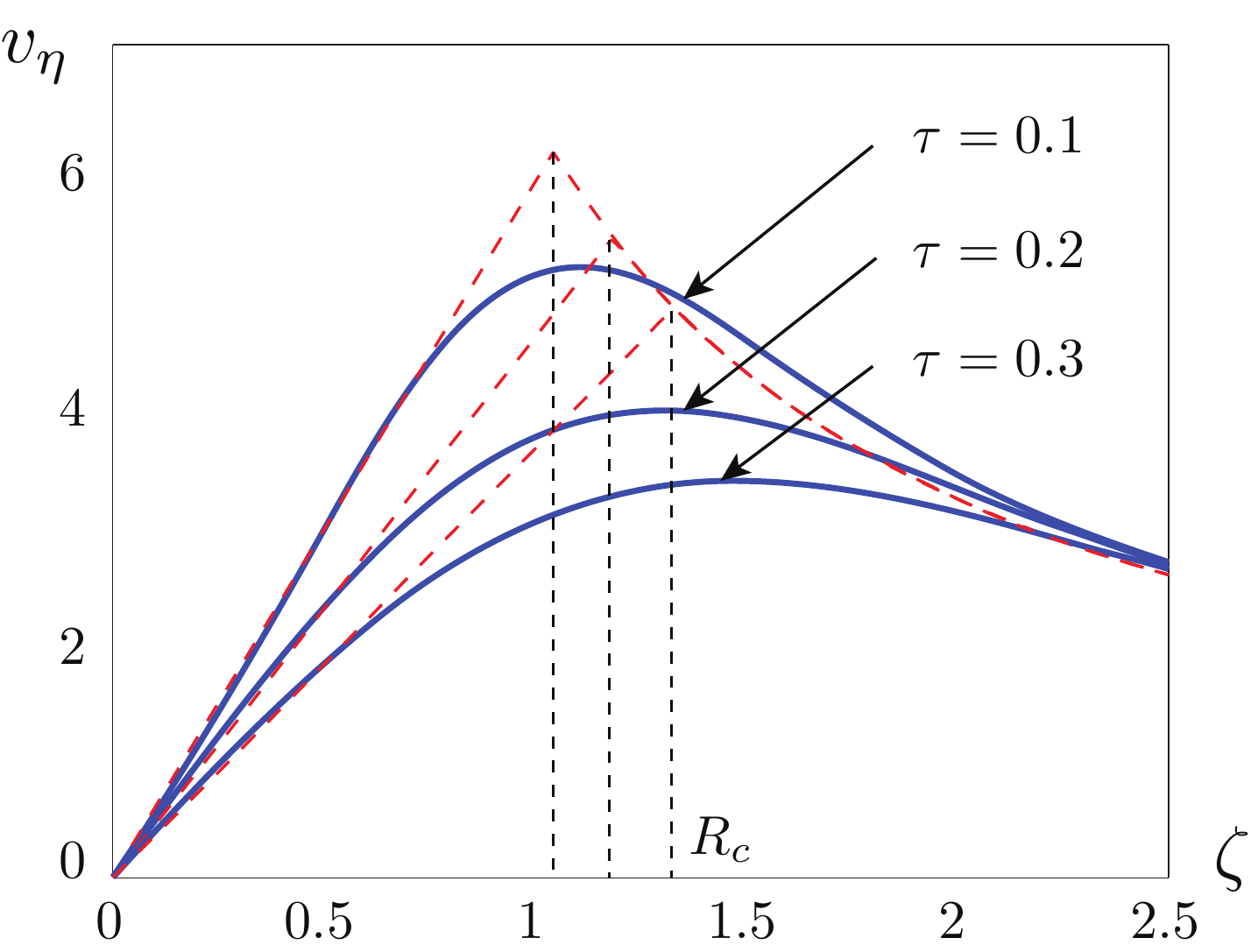} \caption{\footnotesize The velocity field induced by the co-rotating vortex pair becomes analogous to that of a Rankine vortex. The component of velocity $v_{\eta}$ along the $\zeta$ axis is depicted at various instants (solid). The velocity of a Rankine vortex with vorticity $2\Gamma$ and time-dependent core are superimposed (dashed). The velocity field is similar to a rigid rotation close to the origin and an inverse decay at larger distance from the origin.} \label{fig:rankine} 
\end{figure}

\begin{figure}
	[!tb] 
	\begin{center}
		\includegraphics[width=0.85 
		\textwidth]{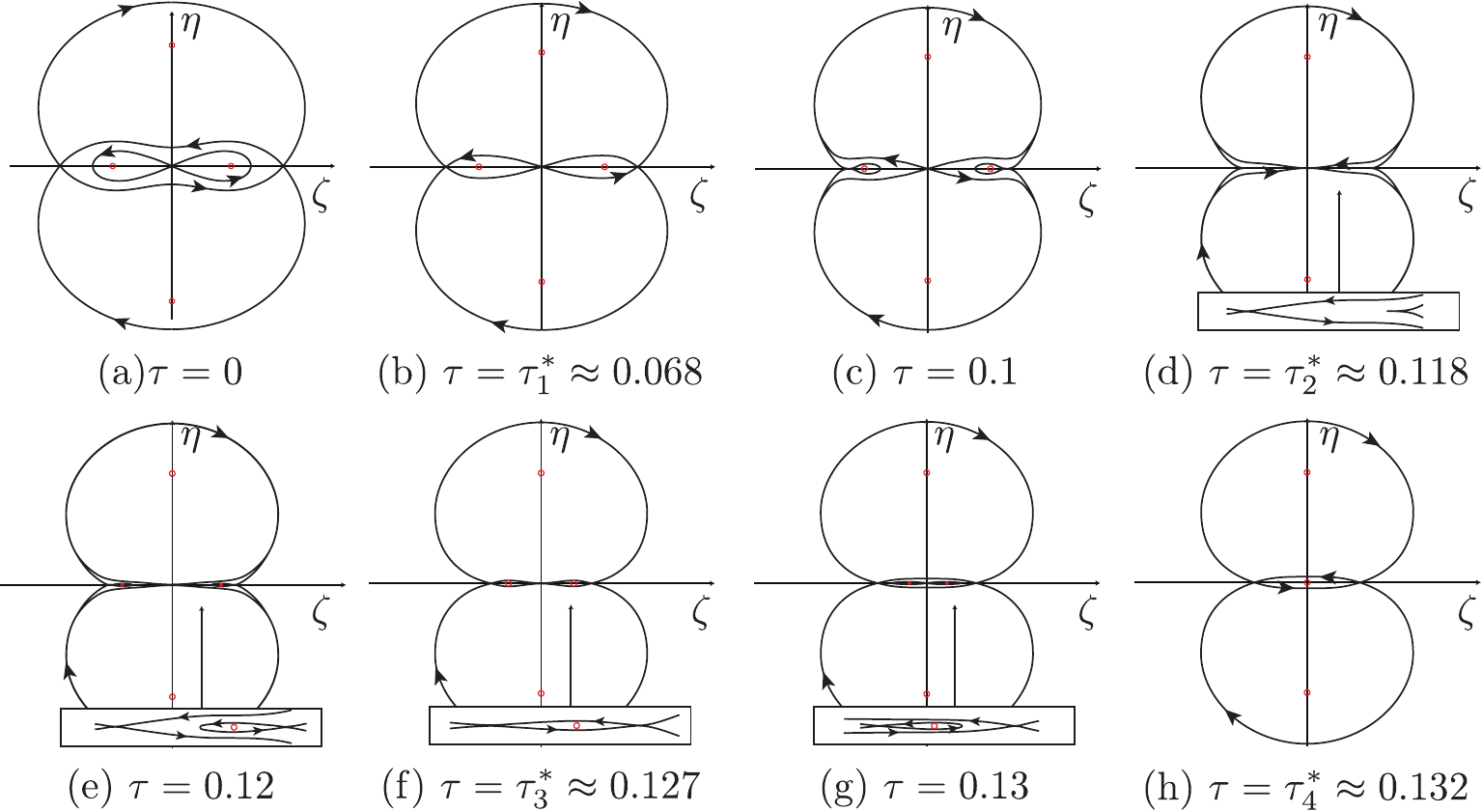} \caption {\footnotesize Evolution of separatrices associated with the relative velocity field of the non-symmetric co-rotating pair in rotating frame. The arrows on separatrices show velocity direction. Hyperbolic points are represented by cross-sections of separatrices. Elliptic points are small circles. $Re = 1000$.} \label{fig:symmseparatrix} 
	\end{center}
\end{figure}

As first pointed out by Dritschel~\cite{Dr1985a} and consequently confirmed by others, the evolution of vortices in a co-rotating reference frame is the key to the understanding of the problem. Traditionally, due to the complexity of its nonlinear nature, finding the rotation rate of the co-rotating frame has been a challenge in both numerical and experimental studies. In our model, however, there exists an obvious choice of co-rotating frame, that is a frame co-rotating with the vortex pair with the time-dependent rotation rate $\dot{\theta}$. Let $\boldsymbol{\xi} = (\zeta,\eta)$ denote position of a point in this rotating frame. The transformation from the rotating to the inertial frame is given by 
\begin{equation}
	{\mathbf{z}} = R \,{\boldsymbol{\xi}}\,, \quad R = \left[ 
	\begin{array}{cc}
		\cos\theta & -\sin\theta\\
		\sin\theta & \cos\theta 
	\end{array}
	\right]\,. \label{eq:positiontransform} 
\end{equation}
\paragraph{Symmetric case.} For the symmetric case, the positions of vortex centers in this rotating frame remain at $\boldsymbol{\xi}_L = (-b_{0}/2,0)$ and $\boldsymbol{\xi}_R = (b_{0}/2,0)$ for all time. As time evolves, the vorticity field, initially concentrated at $\boldsymbol{\xi}_L$ and $\boldsymbol{\xi}_R$, begins to spread spatially, inducing a velocity field similar to that of a {\em Rankine vortex} centered at the origin with a time-dependent Rankine core $R_{c}$. The reader is reminded that the velocity field of a Rankine vortex is such that for the region inside the core, the velocity field corresponds to a {\em rigid rotation}, while for the region outside the core, the velocity field decays proportionally to the inverse of the distance. In Figures~\ref{fig:rankine}, we superimpose the velocity field of a Rankine vortex on the vertical component of the velocity field induced by the viscously evolving co-rotating pair along the horizontal axis at three different instances. The core size $R_{c}$ is obtained by equating $2\Gamma/2\pi R_{c}^{2} = \dot{\theta}$, and it increases in time. As the distance from the origin increases, the velocity field of the co-rotating pair structure decays analogously to the inverse decay with vorticity $2\Gamma$. Since we are interested in the dynamics close to the origin, motivated by this analogy with the Rankine vortex, we examine the time evolution of the {\em relative velocity field} obtained by subtracting a rigid body rotation from the absolute velocity field expressed in the rotating frame, namely 
\begin{equation}
	\dot{\boldsymbol{\xi}} = R^{T} \dot{\mathbf{z}} - R^{T}\dot{R}R^{T}\mathbf{z} = R^{T} \dot{\mathbf{z}} - \dot{\theta} \boldsymbol{\xi}^\perp\,, \label{eq:relvel} 
\end{equation}
where $\boldsymbol{\xi}^\perp = (-\eta,\zeta)$. Note that the second term corresponds to the rigid body rotation of the rate $\dot{\theta}$. Hence the relative velocity is the difference between the rotated absolute velocity field and a rigid body rotation, i.e. the difference between the solid line and the dashed line close to the origin in Figure~\ref{fig:rankine}. The instantaneous separatrices and stagnation points (elliptic points represented by small circles and hyperbolic points by the intersections of the separatrices) associated with~\eqref{eq:relvel} are depicted in Figure~\ref{fig:symmseparatrix}. Initially, one finds a total of 7 stagnation points: a hyperbolic point at the origin $(0,0)$, a pair of elliptic points on the horizontal axis $(\pm b_{0}/2, 0)$, a pair of hyperbolic points on the horizontal axis denoted by $(\pm \xi^{*},0)$ (initially $\xi^{*}(0) = b_{0}\sqrt{5}/2$), and a pair of elliptic points on the vertical axis $(0,\pm b_{0}\sqrt{3}/2)$ which correspond to the so-called ``ghost'' vortices. The elliptic points on the vertical axis remain stationary and of elliptic character for all time. The origin and the pair $(\pm b_{0}/2, 0)$ remain fixed points for all time but their characters change when the pair $(\pm \xi^{*},0)$ passes them as it approaches the origin. Consequently, the topology of instantaneous separatrices changes in time, as depicted in Figure~\ref{fig:symmseparatrix}. The topological bifurcation times $\tau_{2}^{*}\approx 0.118$ and $\tau_{4}^{*}\approx 0.132$ correspond to when $(\pm \xi^{*},0)$ collide with $(\pm b_{0}/2, 0)$ and the origin, respectively, while $\tau_{1}^{*}\approx 0.068$ and $\tau_{3}^{*} \approx 0.127$ correspond to the collapse of separatrices in this process. After $\tau_{4}^{*}$, the topology of the separatrices remains the same. The sequence of topological bifurcations is summarized in Figure~\ref{fig:symmtopology}. The topology of the separatrices at the bifurcation values are depicted in boxes. 
\begin{figure}
	[!tb] 
	\begin{center}
		\includegraphics[width=0.85 
		\textwidth]{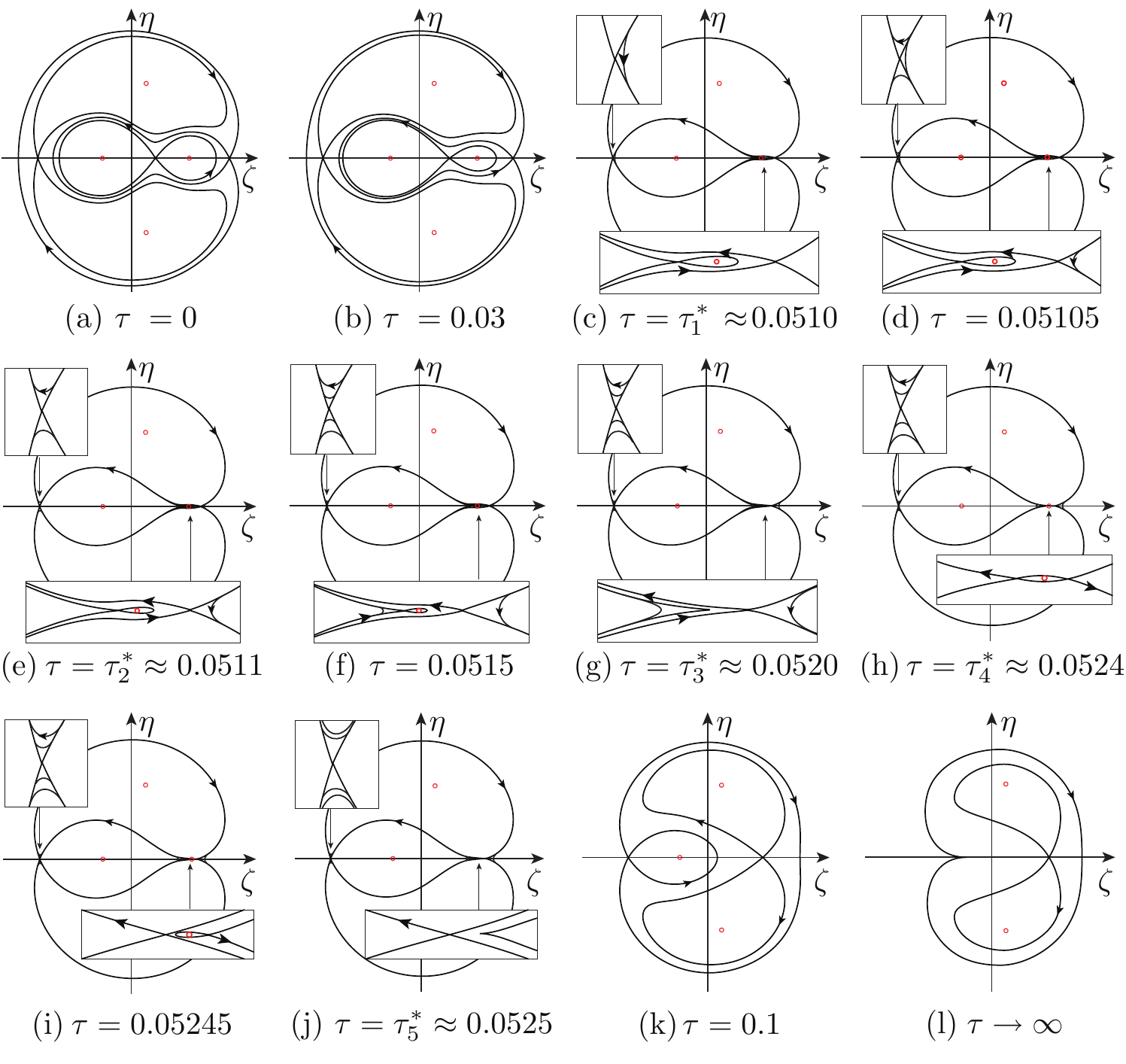} \caption {\footnotesize Evolution of fixed points and separatrices associated with the relative velocity field for the symmetric co-rotating pair in rotating frame. The arrows on separatrices show velocity directions. Hyperbolic points are represented by cross-sections of separatrices. Elliptic points are small circles. $Re = 1000$.} \label{fig:asymmseparatrix} 
	\end{center}
\end{figure}

\paragraph{Asymmetric case.} Following a similar procedure, we examine the bifurcation sequence of the relative velocity field of the asymmetric case. Initially, 7 stagnation points exist: 3 hyperbolic points and 4 elliptic points. Notice since the flow field is not symmetric about the vertical axis (but symmetric about the horizontal axis), the pair of elliptic points is not located on either axes (the other 5 stagnation points are on the horizontal axis). As time evolves, the $1^{\text{st}}$, $2^{\text{nd}}$ and $4^{\text{th}}$ bifurcations are due to collapsing of separatrices; while the $3^{\text{rd}}$ and $5^{\text{th}}$ bifurcations are due to the colliding of instantaneous stagnation points; these 5 bifurcations occur within a very short period of time ($\tau_{1}^{*} \approx 0.0510, \tau_{2}^{*} \approx 0.0511, \tau_{3}^{*} \approx 0.0520, \tau_{4}^{*} \approx 0.0524$ and $\tau_{5}^{*} \approx 0.0525$). The topology of separatrices remains the same with the second to last state after the $5^{\text{th}}$ bifurcation time. The last bifurcation occurs only when $\tau \rightarrow \infty$, though the meaning of instantaneous stagnation points and separatrices diminishes now since the velocity field is zero everywhere. The topological sequence of bifurcations is summarized in Figure~\ref{fig:asymmtopology}.

\subsection{Passive tracer evolution}\label{sec:tracer}

\begin{figure}
	[!tb] 
	\begin{center}
		\includegraphics[width=0.62\textwidth]{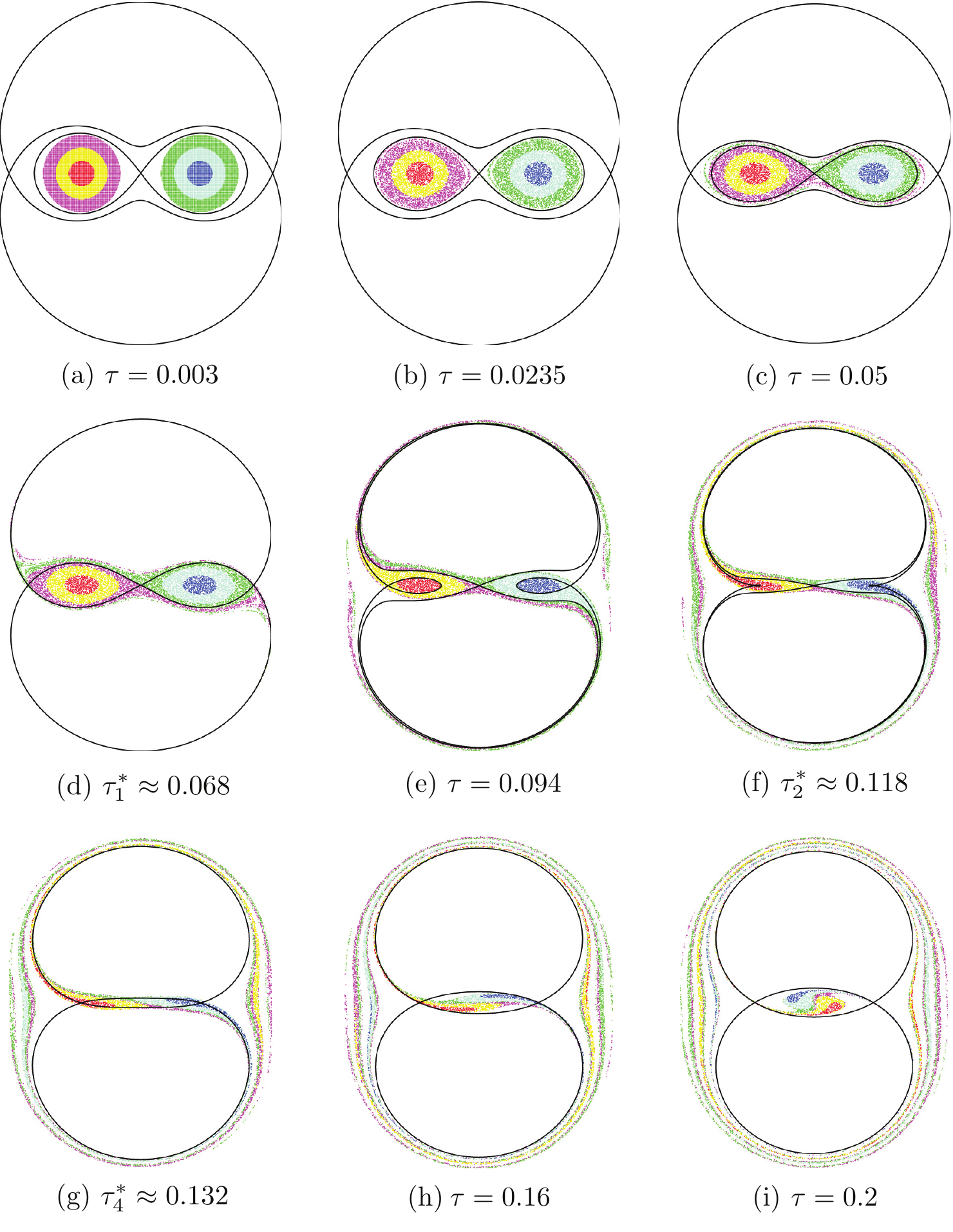} \caption {\footnotesize Passive tracer evolution for the symmetric case.} \label{fig:symmtracer} 
	\end{center}
\end{figure}

As we mentioned in our previous works~\cite{JiKaNe2010a, Ji2011a}, passive tracer evolution in the multi-Gaussian model reveals useful information about the vorticity evolution in the Navier-Stokes equations. This is because weaker vorticity tends to be ``passively" advected by stronger more concentrated vorticity in the full equations of motion (i.e. {\em one-way} coupling of the vorticity field). Interestingly, the evolution of passive tracers by the fluid velocity field is tightly linked to the streamline evolution of the relative velocity field shown in Section~\ref{sec:relative}. This point will be discussed further in Section~\ref{sec:discussion}. Since at time $\tau = 0$ the vorticity is concentrated at two points, we let the vortex cores diffuse for a short amount of time $\tau_{0}$, then seed the flow with passive tracers of six different colors as shown in Figure~\ref{fig:symmtracer} to distinguish tracers within the left (red) and right (blue) cores of radii $a_{0} = \sqrt{4\tau_{0}}$, left (yellow) and right (teal) rings of inner radii $a_{0}$ and outer radii $2a_{0}$, and left (purple) and right (green) rings of inner radii $2a_{0}$ and outer radii $3a_{0}$. Since $a_{0}$ is the core, i.e. the standard deviation of a Gaussian function, the $3a_{0}$ range accounts for more than $99\%$ of the total initial vorticity distribution, known as the \emph{3-sigma rule}. We let the passive tracers be advected by the fluid velocity field given in~\eqref{eq:velmgm}. The snapshots of tracer evolution are rotated into the rotating frame for easier comparison. The reader is reminded that although the general trend of the evolution should only be dictated by the bifurcations of the relative velocity field, the visual details of passive tracer plots will also depend on the way the tracers are seeded, which should be kept in mind when comparing with dye visualization in previous experimental results. One is referred to Figure~13 in~\cite{JiKaNe2010a} for the result of a different way of seeding particles for the collinear state.
\begin{figure}
	[!t] 
	\begin{center}
		\includegraphics[width=0.3
		\textwidth]{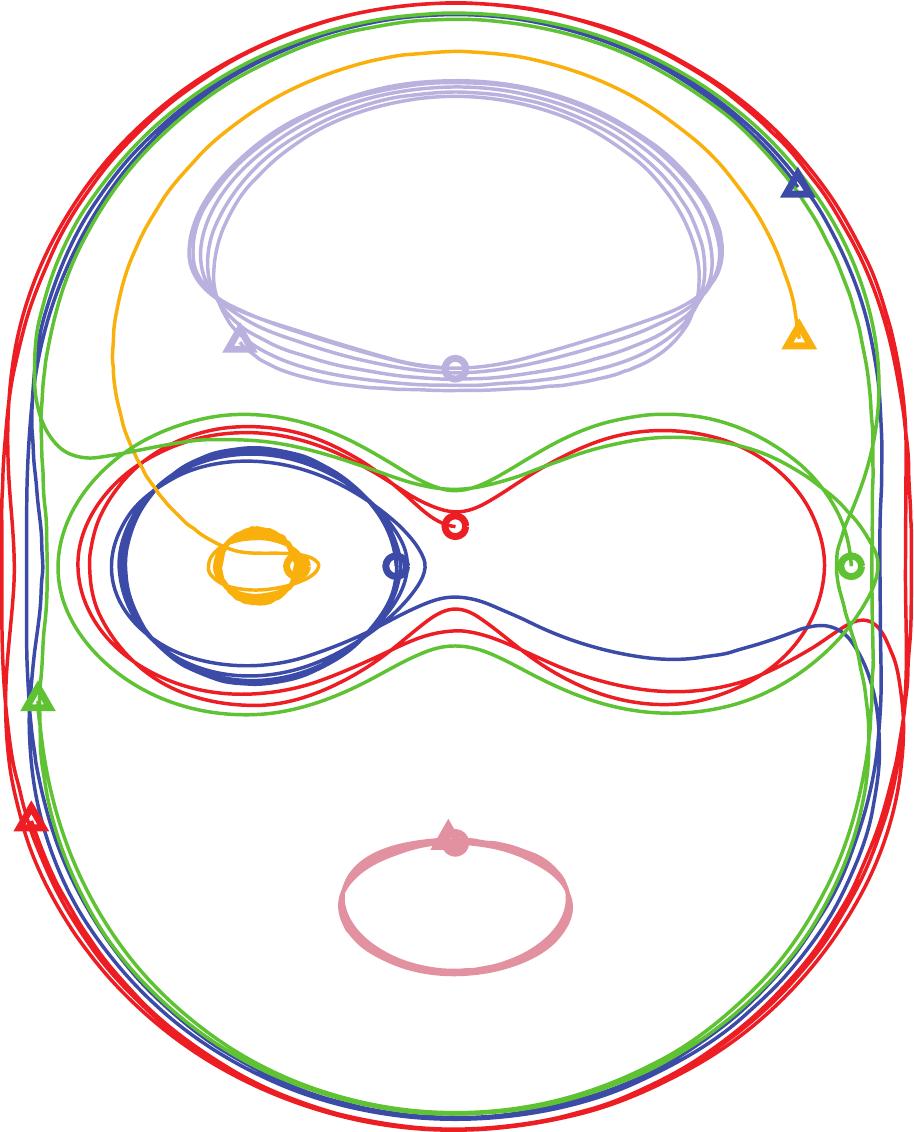} \caption {\footnotesize Trajectories of passive tracers for the symmetric case. Their initial positions at $\tau = 0$ are represented by $\bigcirc$: $(0\,,\,0.1)$, $(0\,,\,0.5)$, $(0\,,\,-0.7)$, $(-0.15\,,\,0)$, $(-0.4\,,\,1)$ and $(1\,,\,0)$; and their final position at $\tau = 0.2$ are marked with $\bigtriangleup$.} \label{fig:path} 
	\end{center}
\end{figure}
\begin{figure}
	[!b] 
	\begin{center}
		\includegraphics[width=0.62 
		\textwidth]{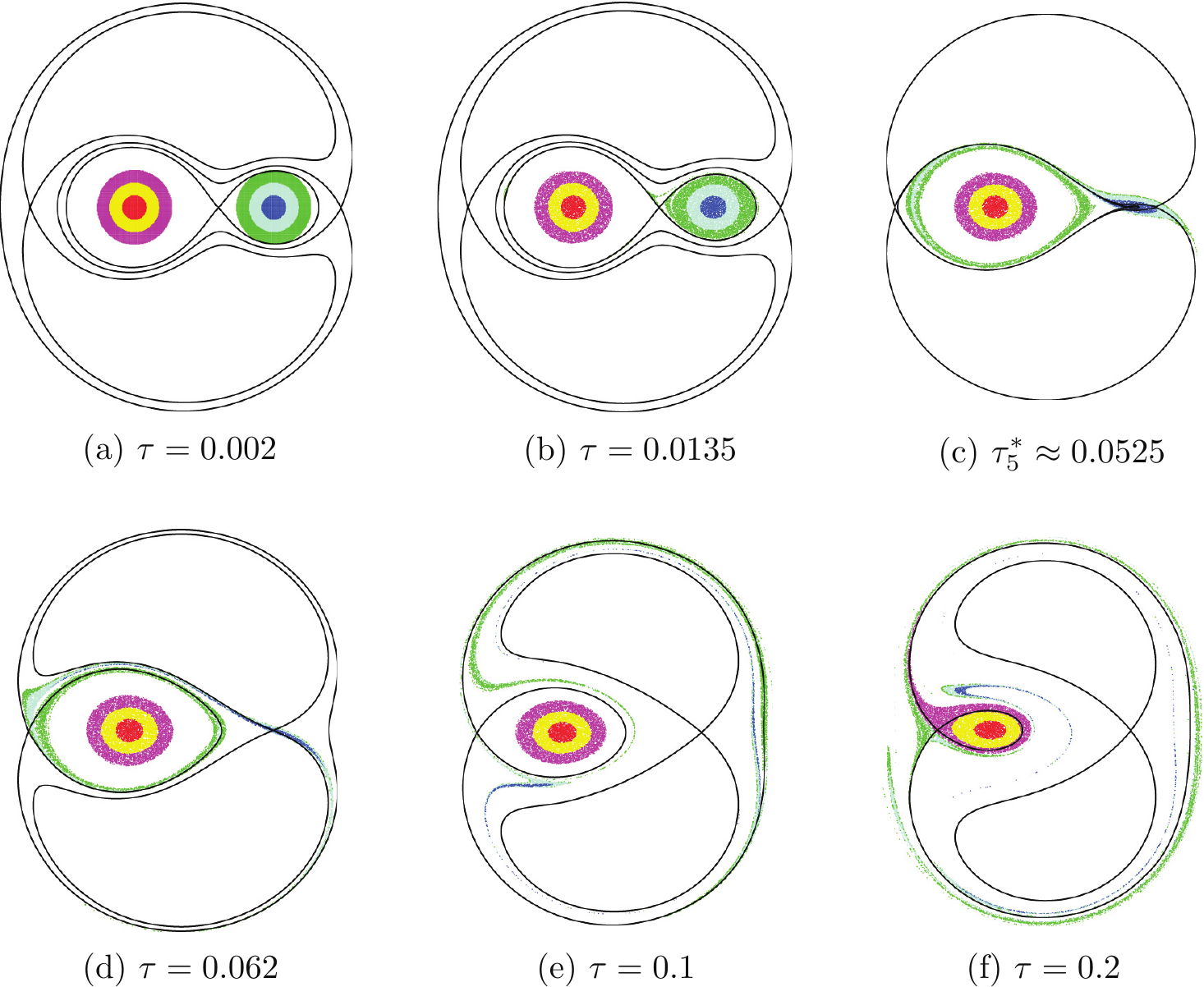} \caption {\footnotesize Passive tracer evolution for the asymmetric case.} \label{fig:asymmtracer} 
	\end{center}
\end{figure}

\paragraph{Symmetric case.} We let $\tau_{0} = 0.003$, hence $a_{0} = \sqrt{4\tau_{0}} \approx 0.110$. The initial state is shown in Figure~\ref{fig:symmtracer}(a). As time evolves, the tracers first rotate around the two elliptic points which coincide with the vortex centers. Before $\tau = 0.0235$, they are confined within the inner separatrix (the region bounded by the inner separatrix is referred to as the \emph{inner cores} hereafter), see Figure~\ref{fig:symmtracer}(b). After $\tau = 0.0235$, some of the outer tracers (purple and green) start to ``leak'' into the band between inner separatrix and middle separatrix (referred to as the \emph{exchange band}), and they mix while circulating in the band around the inner cores. As the middle separatrix continues to shrink, some of the outer tracers will consequently leak outside the exchange band when they encounter the hyperbolic points (along the unstable manifolds associated with the hyperbolic points). At the first bifurcation time $\tau_{1}^{*} \approx 0.068$, the exchange band disappears, the inner tracers still remain in the inner cores, some of the outer tracers are in the inner cores while the rest of the outer tracers now erode in the regions bounded by the outer separatrix and the middle separatrix (referred to as the \emph{outer recirculation area}) which start to form filaments, see Figure~\ref{fig:symmtracer}(d). Since the separatrices are time dependent Eulerian streamlines, the passive tracers will not faithfully follow the separatrices. Therefore when the tracers travel along the outer separatrix and recirculate close to the opposite hyperbolic point, they will ``escape'' to the region outside the outer separatrix (referred to as the \emph{outside area}), see Figure~\ref{fig:symmtracer}(e). In the mean time, some of the middle (yellow and teal) and later inner (red and blue) tracers encounter the hyperbolic points and enter the outer recirculation area, they mix with the outer tracers while traveling close to the outer separatrix. At the second bifurcation time $\tau_{2}^{*} \approx 0.118$, the moving hyperbolic points collide with the elliptic points, and most of the inner tracers still remain close to the now hyperbolic points $(\pm b_{0}/2,0)$, see Figure~\ref{fig:symmtracer}(f). As the separatrices continue to evolve, the tracers continue to mix in the outside area, but closer to the origin the six colored tracers are separated by clear boundaries. After the last bifurcation time $\tau_{4}^{*} \approx 0.132$, the topology of the separatrices remain the same, and the shape of the streamlines will not vary much (although the magnitude of the velocity will continue to decay). Hence, after $\tau_{4}^{*} \approx 0.132$, the tracers confined within the inner separatrix (which include all six colors) will mostly remain inside, and rest of the tracers will circulate around the outer separatrix in the outside area. As shown in Figure~\ref{fig:symmtracer}(h) and (i), due to the difference in the magnitude of velocities inside the inner separatrix, the boundaries between the inside six colored tracers will start to stretch, causing the tracers to mix in layers. Note the bifurcation time for the absolute velocity field $\tau_b \approx 0.050$ is not a significant time, but the time the vorticity peaks merge $\tau = 0.125$ is inline with the time that the inner tracers start to merge. Another interesting fact is that the ``onset'' time for the tracers to leak into the exchange band $\tau = 0.0235$ corresponds to the core size $a = 0.3066$, which is very close to the critical condition of the beginning of merging obtained by the previous studies~\cite{MeLe2001a, CeWi2003a}. The passive tracer result closely resembles the previous Navier-Stokes simulations and experimental results, for a comparison, the reader is referred to Figure~14 in~\cite{CeWi2003a}, Figure~4 in~\cite{MeLeLe2005a} and Figure~1 of~\cite{BrNo2007a}, just to name a few. The trajectories of several tracers in the vicinity of stagnation points are shown in Figure~\ref{fig:path}. The trajectories resemble streamlines in the rotating frame, yet they are not equivalent since the velocities of the tracers only follow the \textit{instantaneous} streamlines, which are time dependent. This will be discussed further in Section~\ref{sec:discussion}.

\paragraph{Asymmetric case.} We let $\tau_{0} = 0.002$, hence $a_{0} = \sqrt{4\tau_{0}} \approx 0.089$. The initial state is depicted in Figure~\ref{fig:asymmtracer}(a). As time evolves, the inner separatrix shrinks and the green tracers leak into the exchange band and circulate around the inner cores, see Figure~\ref{fig:asymmtracer}(b). As the inner separatrix continue to shrink, all of the green tracers will enter the exchange band, followed by all the teal tracers and some of the blue tracers, while the left tracers still remain inside the inner core and rotate around the left elliptic point. This is because in this case, the right ring of the inner separatrix shrinks, but the left ring actually expands. As the right hyperbolic point continue to approach the right elliptic point, some of the tracers leak along the unstable manifold. As mentioned before, the 5 bifurcation times occur within a short period of time, hence the positions of the passive tracers will not change much during the bifurcation times. However, the last bifurcation time $\tau_5^*$ marks the end of the state that the tracers ``hover'' around two elliptic points, and all tracers will start to rotate around the remaining left vortex from now on. As shown in Figure~\ref{fig:asymmtracer}(c), the exchange band disappears, all the left tracers remain in the inner core, while some of the right tracers circulate and the others leak to the outer recirculation area or the outer area. In Figure~\ref{fig:asymmtracer}(d) and (e), the blue tracers continue to escape along the unstable manifolds, and the right tracers erode into the band between the outer and inner separatrices. As the left hyperbolic point continues to approach the left elliptic point, the left ring of the inner separatrix continues to shrink, and the purple tracers start to leak outside the left ring, see Figure~\ref{fig:asymmtracer}(f). Since the left ring will shrink to zero only when $\tau \rightarrow \infty$, it will take infinite amount of time for the red tracers to completely leak along the unstable manifold associated with the left hyperbolic point. One can see remarkable resemblance between this result and the dye evolution in~\cite{TrFuHe2005a} (their Figure 15) and vorticity contour in~\cite{BrNo2010a} (their Figure 1 \& 4).

\section{Discussions and conclusions}\label{sec:discussion}

There are three key features in the analysis of the core-growth model we highlight here: 
\begin{enumerate}
	\item The streamline description of a relative velocity field (velocity field minus a rigid body rotation) and their topological bifurcations analyzed in an appropriately chosen time-dependent rotating frame; 
	\item The passive tracer field description in this same rotating frame; 
	\item The identification of the passive field as a {\em one-way coupled} model of vorticity evolution giving insights into the dynamics of a weak vorticity field (the outer region of a decaying Gaussian core) advected by the more strongly concentrated vorticity field emerging from the concentrated initial conditions. 
\end{enumerate}

\begin{figure}
	[!t] 
	\begin{center}
		\includegraphics[width=0.9 
		\textwidth]{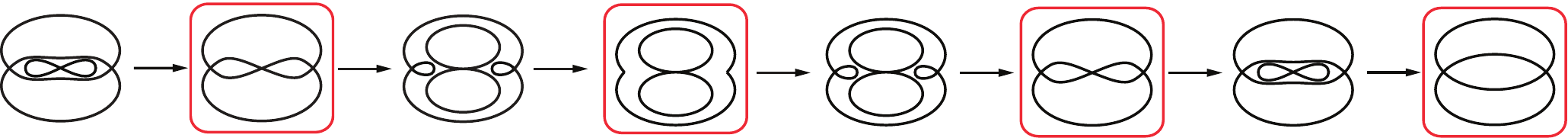} \caption {\footnotesize Homotopic equivalences of the separatrices for symmetric co-rotating pair. Bifurcation states are depicted in boxes.} \label{fig:symmtopology} 
	\end{center}
\end{figure}
\begin{figure}
	[!t] 
	\begin{center}
		\includegraphics[width=0.75 
		\textwidth]{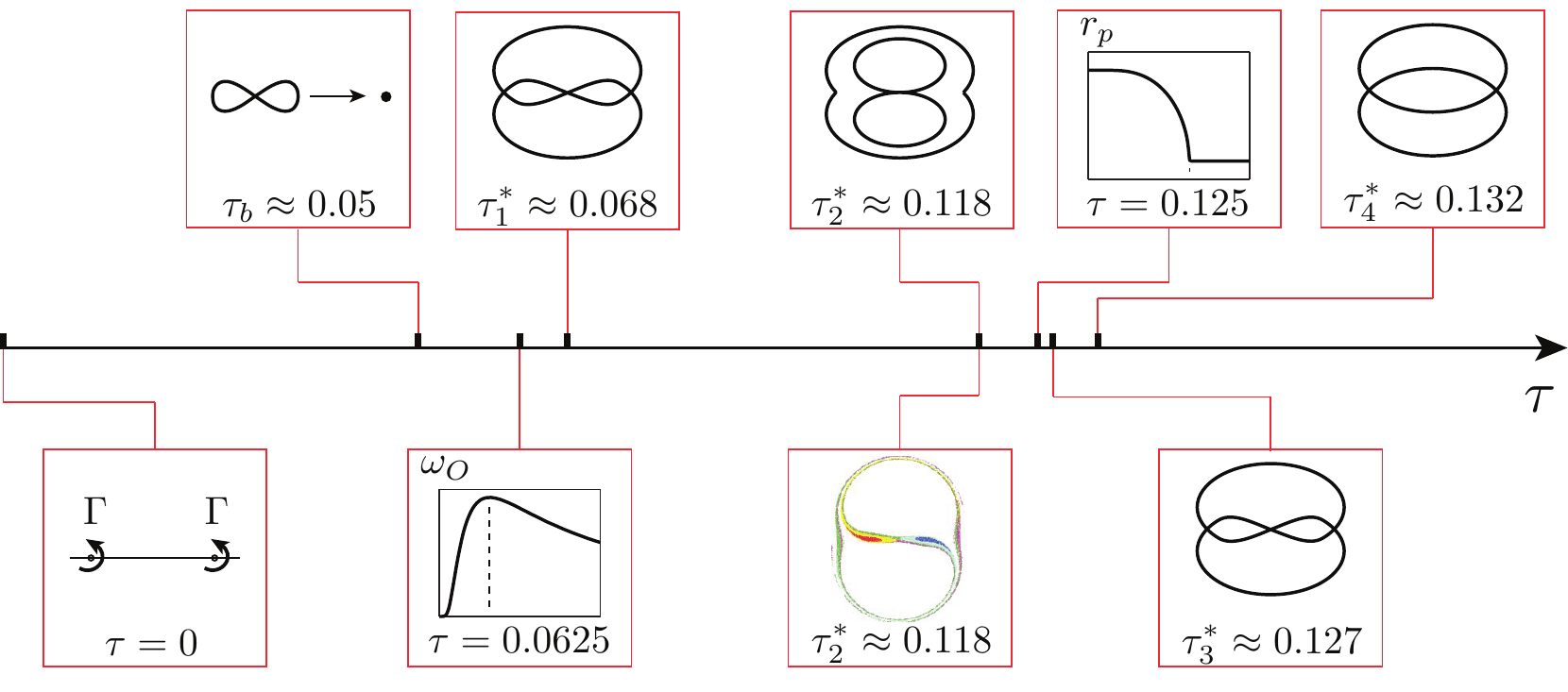} \caption {\footnotesize Timeline of important events for the symmetric case.} \label{fig:symmtimeline} 
	\end{center}
\end{figure}

The description offered by the model reveals much more detailed and delicate structure than could easily be pinpointed by careful experiments addressing the same questions, such as \cite{CeWi2003a}. Their experiment, however, clearly reveals the importance of the {\em two-way coupling} of the vorticity and velocity fields in the convective stage, as the vorticity is squeezed out of the inner zones shown in our Figure \ref{fig:symmtracer}, with respect to the generation of a quadrupole structure which is the key mechanism responsible for pushing the two vortex centers towards each other. This feature cannot be captured in our {\em one-way coupled} description. The general monotonically decreasing trend of the distance between vortex centers as shown in our Figure~\ref{fig:symmomega}(b) is consistent with the general direction. One can compare the separation plot with the Figure 5(a) in the DNS study in~\cite{BrNo2007a} and see good qualitative agreement. From the value $\tau = 0.125$ which is the time two peaks merge in the symmetric case, one can readily calculate that $t/Re \approx 0.0063$. Though this value does not quantitatively agree with the results obtained by experiments (the starting core sizes in experiments are generally around $0.15 b_0$, which contributes to the complexity as well), e.g. on average 0.0011 for $Re = 87, 224, 530$ in~\cite{CeWi2003a}, and 0.0015 for $Re = 742$, 0.0012 for $Re = 1506$, 0.0011 for $Re = 0.0011$ in~\cite{MeLeLe2005a}, one can see that multi-Gaussian model predicts merging time (if defined as the merging of vorticity peaks) roughly 4$\sim$5 times larger, and this difference ratio is consistent with the comparison between the bifurcation time for the model and DNS in our previous work~\cite{JiKaNe2010a}. We believe the difference will decrease once the quadrupole is added to the model.

The rotating frame and topology of streamline patterns is essential to the understanding of viscous vorticity evolution. In the multi-Gaussian model, the angular velocity of the structure $\dot{\theta}$ (which will eventually decrease to zero due to the presence of viscosity) is a natural choice of rotation rate, and the analogy between the velocity field of the vortex pair and that of a Rankine vortex justifies the physical meaning of subtracting a rigid body rotation from the (rotated) absolute velocity field. Indeed, although the passive tracer evolution is a result from integrating the absolute velocity field, it is obvious that the relative velocity fields (Figure~\ref{fig:symmseparatrix} and~\ref{fig:asymmseparatrix}) are much more relevant to the tracer evolution than the absolute velocity fields (Figure~\ref{fig:symminertial} and~\ref{fig:asymminertial}). This is evident from comparison between the passive tracer trajectories (Figure~\ref{fig:path}) and rotating frame streamlines (Figure~\ref{fig:symmseparatrix}) and inertial frame streamlines (Figure~\ref{fig:symminertial}). The reason was essentially noted by Velasco Fuentes~\cite{Fu2005a} --- the Lagrangian dynamics of the passive tracer trajectories do not always follow the \emph{time varying} Eulerian streamlines. In steady flows, both geometries are identical. The slower the Eulerian velocity field is varying, the less difference between the two~\cite{HaPo1998a}. In this case, the absolute velocity field is a fast varying Eulerian field, which is both rotating and evolving. The relative velocity field is a much slower varying Eulerian field since it is a result of subtracting the fast varying part (rigid body rotation) off the absolute field.

\begin{figure}
	[!t] 
	\begin{center}
		\includegraphics[width=0.85 
		\textwidth]{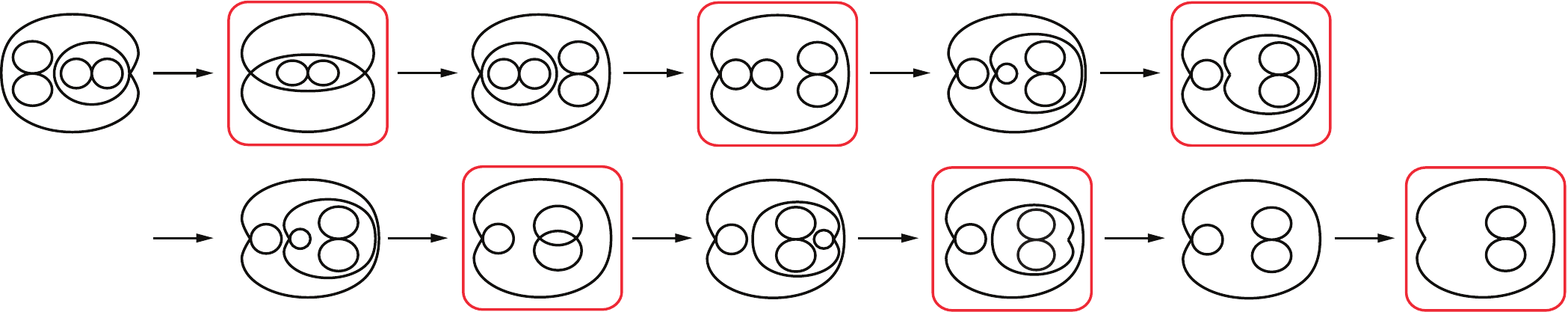} \caption {\footnotesize Homotopic equivalences of the separatrices for asymmetric co-rotating pair. Bifurcation states are depicted in boxes.} \label{fig:asymmtopology} 
	\end{center}
\end{figure}
\begin{figure}
	[!t] 
	\begin{center}
		\includegraphics[width=0.65 
		\textwidth]{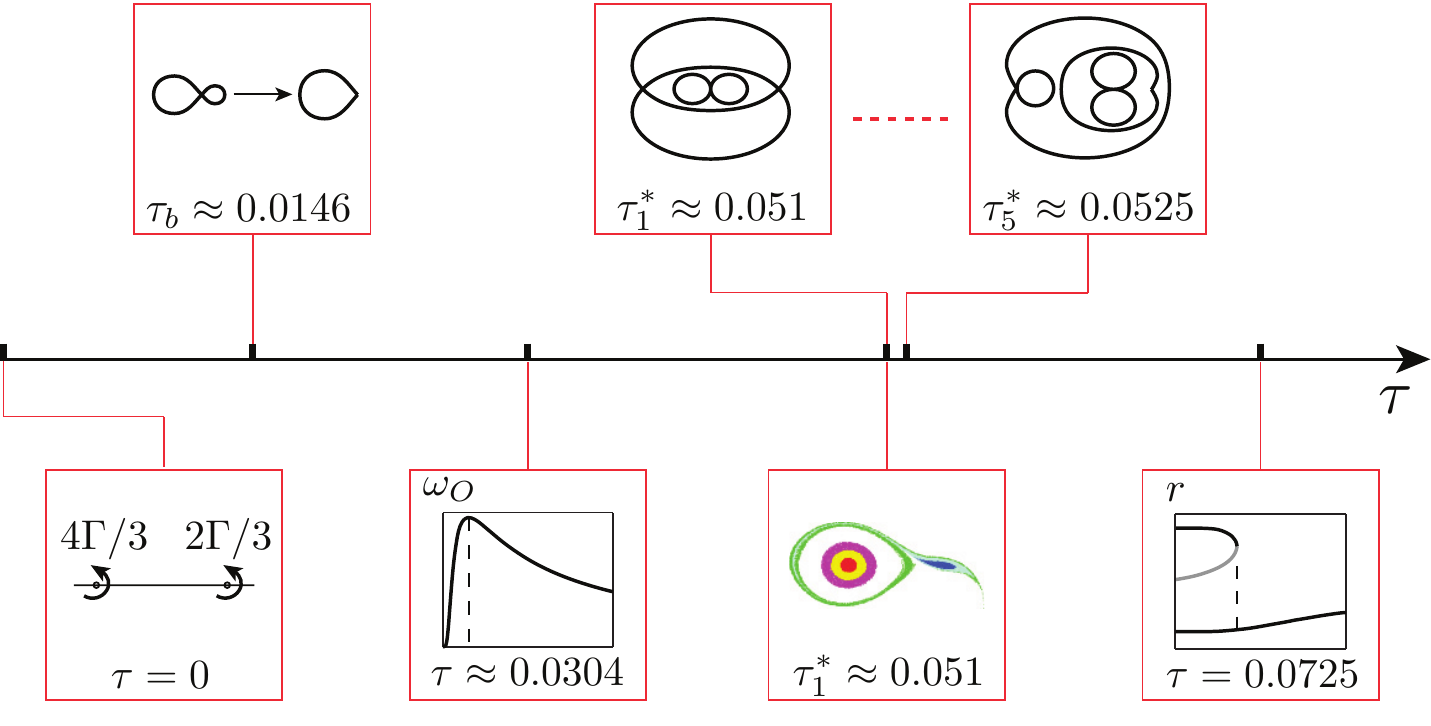} \caption {\footnotesize Timeline of important events for the asymmetric case.} \label{fig:asymmtimeline} 
	\end{center}
\end{figure}

By using the multi-Gaussian model, we are able to analyze the topological evolution and bifurcation sequence of the relative field separatrices, and obtain the topological bifurcation times. The results can be partially verified in comparison to the experimental results obtained by previous studies of the symmetric case. The best example is Figure 13 in Cerretelli \& Williamson~\cite{CeWi2003a}, where they obtained separatrices topologically identical to Figure~\ref{fig:symmseparatrix}(a) and (h), but the intermediate topological states were not shown. Indeed, we have not proven that they exist for the full Navier-Stokes equations, only for the multi-Gaussian model. For the asymmetric case, to our knowledge, no previous research has shown separatrices in the rotating frame other than the initial state. We believe it is essential to understand the topological evolution of the streamlines, which dictate the merging process, and this work can be viewed as a first step towards unveiling the Navier-Stokes evolution of streamlines in the merging process. We present the complete sequence in Figure~\ref{fig:symmseparatrix} for the symmetric case and Figure~\ref{fig:asymmseparatrix} for the asymmetric case, and one can see that the bifurcations are rich and not at all trivial. The timeline of all important events based on the multi-Gaussian model are plotted in Figure~\ref{fig:symmtimeline} for the symmetric case and in Figure~\ref{fig:asymmtimeline} for the asymmetric case.

The path is now clear for a corrected low-dimensional core-growth based model of vortex merger. One needs to augment the basic description in this paper with the development and growth of an appropriately chosen quadrupole structure surrounding the co-rotating state, chosen with time dependent vorticity strengths which develop as the primary vorticity from the original core is squeezed out of the inner core region. It is interesting to note that some of these multi-pole features are systematically used in the work of~\cite{UmWaBa2012a}. Implementing these key features in the context of a low-dimensional model of merger will be the focus of a separate paper. 

\section*{Acknowledgments} The work of FJ and EK was partially supported by the National Science Foundation through the CAREER award CMMI 06-44925 and the grant CMMI 07-57092. Part of the work carried out was supported by the National Science Foundation Grant to PKN under NSF-DMS0804629.


\end{document}